\documentstyle [prl,aps]{revtex}
\newcommand{\be}{\begin{equation}}
\newcommand{\ee}{\end{equation}}
\newcommand{\bea}{\begin{eqnarray}}
\newcommand{\eea}{\end{eqnarray}}

\newcommand{\s}{\sigma}
\newcommand{\rF}{\mbox{F}}

\newcommand{\rd}{\mbox{d}}
\newcommand{\ri}{\mbox{i}}

\begin {document}
\input{psfig}
\bibliographystyle {plain}

\title{\bf Coherence-Incoherence Transition Between Fermi
and Luttinger liquids}
\author {A. M. Tsvelik}
\maketitle
\begin {verse}
$Department~ of~ Physics,~ University~ of~ Oxford,~ 1~ Keble~ Road,$
\\
$Oxford,~OX1~ 3NP,~ UK$\\
\end{verse}
\begin{abstract}
 We consider conditions for existence  of fermionic quasiparticles in
a strongly anisotropic quasi-one-dimensional metal. The adopted model
is  a
model of chains of spin-1/2 Luttinger liquid coupled 
by  small interchain hopping.  It 
is shown that the Luttinger liquid becomes more stable with decrease
of the lattice coordination number. It is also 
shown that with a significant spin-charge
separation some   Luttinger liquid
features may persist in the Fermi liquid phase.  
\end{abstract}

PACS numbers: 71.27.+a, 72.10.-d
\sloppy
\par

 There are 
quasi-one-dimensional materials which, despite being good
conductors, 
display features which are considered  as 
incompatible with the classical Fermi liquid
description. It is suggested that
the adequate description of these materials may be provided by the
Luttinger liquid rather than the Fermi liquid theory.  There are also
materials like (TMTSF)$_2$PF$_6$ which being  Fermi 
liquids under normal conditions
show  manifestly non-Fermi-liquid properties in a magnetic field 
\cite{chaikin}, \cite{dressel}. To give an adequate 
theoretical description of such
systems one must understand a problem of stability of Luttinger liquid
with respect to interchain hopping $t_{\perp}$. 
Despite that this problem has been addressed 
by many distinguished researches, there is still no consensus about the
results. In this letter we argue that the existent results allow  
to establish sufficient conditions for stability of
Luttinger liquids. 

 The most difficult aspect of the problem of stability of Luttinger
liquid  is that the interchain hopping
simultaneously generates effects of different nature. 
Thus there are processes leading  to creation of three
dimensional Fermi surface and processes of multi-particle exchange
leading  to  three dimensional phase transitions. Even if the Fermi
surface is not formed, the interchain exchange will always
destabilize the gapless Luttinger liquid fixed point and create a three
dimensional ground state with a broken symmetry. In this and only this
sense the interchain hopping  is relevant. There is
another question however,  and it is whether the state above
the transition temperature is a normal metal or something else. 
It was rightly
pointed out by Anderson that the earmark  of a Fermi liquid
is a coherent transport in all directions 
(see also the discussions in Refs.\cite{clarke},\cite{cl}). 
Meanwhile, in a one dimensional 
(or ``confined'') Luttinger liquid state
the Drude peak in optical conductivity exists only in the direction 
along the chains. Since the Drude peak in normal metals originates from
existence of a well defined pole in the single electron Green's
functions which is absent in Luttinger liquid, we have to work out
criteria for formation of such a pole.

 To establish sufficient conditions for existence of the single
particle pole one should make the phase transition temperature as low
as possible. This can be achieved on a lattice with a large coordination
number $D >> 1$. This condition will allow 
us to  ignore all feed back
effects coming from electrons returning to the same chain and
generating  effective 
many-body exchange interactions  leading 
to phase transitions at low temperatures. In order to minimize
similarity between the Luttinger liquids on single chains and a Fermi
liquid we shall consider a strong spin-charge separation $v_s/v_c << 1$.

 Spin-1/2 Luttinger liquid can be described as a direct sum of charge
and spin subsystems. The corresponding  Hamiltonians 
describe free bosonic scalar
fields with linear spectra: 
$\epsilon_{c, s}(q) = v_{c, s}|q|$. Except  for  the velocities $v_{c,
s}$ the U(1)$\times$SU(2)-symmetric 
Luttinger Hamiltonian includes only one phenomenological
parameter $K_c$ characterizing scaling dimensions 
in the charge sector. In our discussion we shall treat 
$K_c$ as an 
arbitrary parameter.

 Taking into account only the diagrams which can be cut along a single
hopping integral line as being leading ones in $1/D$,  we get the following
expression for the single electron 
Green's function:
\be
G(\omega, k, {\bf k}) = [G_0^{-1}(\omega, k) - t_{\perp}({\bf
k})]^{-1} \label{interchain}
\ee
Here  $G_0(\omega, k)$ is the Fourier transformation of the single chain 
Green's function of the spin-1/2 Luttinger liquid, ${\bf k}$
represents 
transverse
components of the wave vector and 
\[
t_{\perp}({\bf k}) = t_{\perp}\sum_{i = 1}^D\cos({\bf k}{\bf b}_i)
\]
with ${\bf b}_i$ being lattice vectors. 

 Eq.(1) was studied  by Wen \cite{wen} who used various approximate
expressions for $G_0(\omega, k)$. Wen demonstrated 
that at $T = 0$ the pole appears when $\theta < 1$ with  
\[
\theta = (K_c + K_c^{-1} - 2)/4 
\]
 
 Below we will partially repeat the Wen's arguments and 
derive some new results. The retarded Green's function has 
singularities near right and left
Fermi points $k = \pm k_{\rF}$.  In what
follows we shall concentrate on the vicinity of the right Fermi point
and introduce the wave vector $q = k -  k_{\rF}$.
The calculations of  the single particle Green's function
of the Luttinger liquid carried out in Ref.\cite{voit} have
established that at $T = 0$ its imaginary part (the spectral function
$\rho(\omega, q)$)
is finite in the area represented on Fig. 1. In general we have 
\be
G_0^{(R)}[\omega, k = \pm (q + k_{\rF})] =
\frac{|q/k_{\rF}|^{\theta}}{qv_c}\left[{\cal F}_1(\omega/v_sq) + 
\sin(\pi\theta/2){\cal F}_2(\omega/v_cq)\right]
\ee
The functions ${\cal F}_i$ can be extracted from the 
calculations of $G_0$ performed in Ref. \cite{voit}. For $q > 0$ we
have :
\bea
{\cal F}_1(\omega/v_sq) = (v_c/v_s)^{1 + \theta}\nonumber\\
\times\int_0^{v_c/v_s - 1}
\rd x x^{-1/2}(v_c/v_s - 1 - x)^{- 1/2 - \theta/2}(x + v_c/v_s + 1)^{-
\theta/2}(x + 1 + \omega/v_sq)^{- 1 + \theta} \label{gr1}
\eea
\bea
{\cal F}_2(\omega/v_cq) = \int_0^{\infty} \rd x x^{- \theta/2}(x +
2)^{- 1/2 - \theta/2}(x + 1 + v_s/v_c)^{-1/2}(x + 1 - \omega/v_cq)^{-
1 + \theta} \nonumber\\
+ \int_0^{\infty} \rd x x^{- 1/2 - \theta/2}(x +
2)^{- \theta/2}(x + 1 -  v_s/v_c)^{-1/2}(x + 1 - \omega/v_cq)^{-
1 + \theta} \label{gr2}
\eea
For $\theta < 1$ these integrals converge at small $x$ and the answers
are universal. 

To study  the Fermi surface formation, we need to know $G_0(\omega, q)$ at
$\omega \rightarrow 0$. As follows from Eqs.(\ref{gr1}, \ref{gr2}),
in the area $- v_c q < \omega < v_s q$ ($q >
0$) 
the Green's function is purely real 
and can be expanded in powers of $\omega$.  At $v_s/v_c << 1$ the
leading contribution comes from the ${\cal F}_1$-term:  
\bea
{\cal F}_1(\omega/v_sq) =  a -  b\frac{\omega}{qv_s} + ... \label{answer}
\eea
 
 Substituting (\ref{answer}) into
Eq.(1) we obtain the following expression for the dispersion law:
\be
\epsilon(q, {\bf k}) = \frac{a}{b}v_s\left[- q + av_c^{-1}t_{\perp}({\bf
k})(|q|/k_{\rF})^{\theta}\right] \label{disp}
\ee
(mind that $q$ is counted from the old Fermi vector and therefore $q
\neq 0$ at the new Fermi surface defined by $\epsilon(q_{\rF},{\bf k})
= 0$).  So we see, that as it was originally found by Wen, 
the Fermi surface is formed at
$\theta < 1$ which corresponds to $3 - 2\sqrt 2 < K_c < 3 + 2\sqrt
2$. Large $K_c$ are probably unphysical, but small ones are certainly
not.  The widely
quoted restriction $K_c > 1/2$ is valid only for the Hubbard
model. Small $K_c$'s are ubiquitous in
CDW (charge density waves) materials (see, for example 
the review article \cite{gruner},
having in mind  that in the literature on CDW  
$K_c$ is called
$(m/m^*)^{1/2}$). The reduction of $K_c$ comes mainly from
retardation processes related to 
the  electron-phonon interaction 
\cite{braz},\cite{fukuyama}. The phonon-mediated interaction may also
open a spin gap which would be undesirable for Luttinger liquid.  
However, the  tendency to open a gap may be suppressed by  a  strong
electron-electron repulsion which does not affect  the retardation
processes.

 From Eqs.(1) and (\ref{disp}) we find the expressions for the new Fermi
vector
\be
\delta k_{\rF} = k_{\rF}\left[at_{\perp}({\bf
k})/\epsilon_{\rF}\right]^{1/1 - \theta}
\ee
(here and below $\epsilon_{\rF} = v_ck_{\rF}$) and the quasiparticle residue:
\be
Z \approx \frac{v_s}{v_c}\frac{a^2}{b}\left[at_{\perp}({\bf
k})/\epsilon_{\rF}\right]^{\theta/1 - \theta} \label{Z}
\ee

 As we shall see in a moment, the spin-charge separation plays an
important role in the behavior of $Z$ providing  an additional small
parameter in the region $\theta < 1/2$. 

 From Eq.(\ref{gr1}) we find 
\bea 
a \approx (v_c/v_s)^{1 + \theta}\int_0^{v_c/v_s}
\rd x x^{-1/2}(1 + x)^{- 1 + \theta}(v_c/v_s - x)^{- 1/2 - \theta/2}(x
+ v_c/v_s)^{- \theta/2} \label{A}
\eea
The integral for  $b$ is determined by $x \sim 1$ and can be calculated
analytically:
\bea
b \approx (1 - \theta)(v_c/v_s)^{1/2}B(1/2, 3/2 - \theta) \label{B}
\eea
where $B(x,y)$ is the beta function. The integral (\ref{A}) has different areas 
of convergence depending on
whether $\theta < 1/2$ or $1/2 < \theta < 1$. In the former  case it
converges at $x \sim 1$:
\be
a \approx (v_c/v_s)^{1/2}B(1/2, 1/2 - \theta) \label{A1}
\ee
In the latter case the integral converges at large $x$ such that  one can
neglect a difference between $x$ and $x + 1$ to obtain
\be
a \approx (v_c/v_s)^{\theta}B(\theta - 1/2, 1 - \theta/2)F(\theta/2,
\theta - 1/2, 1/2 + \theta/2; - 1)  \label{A2}
\ee
where $F(a,b,c;x)$ is the hypergeometric function. Substituting Eqs.(\ref{A1}, 
\ref{B}, \ref{A2}) into (\ref{Z}) we get 
\be
Z \sim  \left(\frac{v_s}{v_c}\right)^{\gamma}\left[t_{\perp}({\bf
k})/\epsilon_{\rF}\right]^{\theta/1 - \theta}
\ee
where 
\bea
\gamma = (1 - 2\theta)/2(1 - \theta) > 0,  \:\: (\theta < 1/2)\nonumber\\
\gamma = (3 - \theta)(1 - 2\theta)/2(1 - \theta) < 0\:\: (1 > \theta > 1/2) 
\eea

 Therefore we see that the exponent $\gamma$ changes its sign at
$\theta = 1/2$ which leads to a rather peculiar situation. 
Namely, at $\theta < 1/2$ the single particle pole is formed, but its
residue may be small even at moderately small  $t_\perp/\epsilon_{\rF}$. This
means that the Fermi liquid phase will retain 
many features of the Luttinger liquid (see Fig. 2). So for $\theta < 1/2$ 
the spin-charge separation works in favor of the Luttinger
liquid. For $\theta > 1/2$ the situation is reversed: one needs a  very
high degree of one-dimensionality to get  small $Z$. This apparently
paradoxical behavior has a simple explanation. As follows from
\cite{voit}, at $\theta < 1/2$ the single particle spectral function
is singular at $\omega = v_s q$ and $\omega = v_c q$ (see Fig. 2
(a)). In the presence of hopping these singularities are smeared out,
but the maxima still remain (Fig. 2(b)) 
and therefore, as we have said,  many features
of the system are still Luttinger-liquid-like. At $\theta > 1/2$ these
singularities disappear and the spectral function becomes
featureless. Then when the pole is formed we will have a situation
very much like in a conventional Fermi liquid: a featureless
incoherent background and a sharp coherent pole. 

 Though all the above calculations have been carried out at zero
temperature, we can make certain qualitative estimates for $T \neq
0$. According to the general principle of conformal field theory one
can  obtain expressions for finite temperature thermodynamic 
correlation functions
from the $T = 0$ expressions by replacing powers of 
$(\tau \pm \ri x/v)$ by  the powers of $\sin\{\pi T(\tau \pm \ri
x/v)]\}/\pi T$ ($\tau$ is the Matsubara time). After doing all
necessary analytic continuations one obtains the following general
expression which replaces Eq.(2) at finite $T$: 
\be
G_0^{(R)}[\omega, k = \pm (q + k_{\rF});T] = \frac{1}{T}
(T/\epsilon_{\rF})^{\theta}{\cal F}(\omega/T, v_sq/T) \label{finiteT}
\ee
where ${\cal F}(x, y)$ is such that it is finite at $(x, y)
\rightarrow 0$ and at
$(|\omega, q|) >> T$ reproduces Eq.(2) (an explicit expression for
${\cal F}(x, y)$ will be published elsewhere). This consideration gives us an
estimate of the temperature where a crossover from the Luttinger to Fermi
liquid takes place:
\be
T_{cross} \sim \epsilon_{\rF}\left[a t_{\perp}({\bf
k})/\epsilon_{\rF}\right]^{1/1 - \theta} \label{cross}
\ee

 Recently Clarke and Strong have also 
suggested \cite{cl1} that the region $1/2 <
\theta < 1$ is somewhat special.  It is curious, that 
 the incoherent
transport also  undergoes  change at $\theta > 1/2$ . Namely, (see
Eq.(\ref{conduct})) the incoherent part of the transverse optical
conductivity looses a  characteristic  maximum at $\omega  = 0$.

 In the leading order in $t_{\perp}$  the real part of the transverse
conductivity is given by the following expression 
(see, for example \cite{wen}):
\bea
\s_{\perp}(\omega) 
= \frac{e^2t^2_{\perp}}{2
\pi^2\omega}\int\rd q\rd x[n(x) -
n(\omega + x)]\rho(x, q;T)\rho(\omega + x, q; T) \label{S}
\eea
where $n(x)$ is the Fermi distribution function and $\rho(x, q;T)$ is 
the finite temperature spectral
function. Using the scaling
form (\ref{finiteT}) we get 
\bea
\s_{\perp}(\omega, T) = T^{- 1 + 2\theta}{\cal S}(\omega/T) \label{conduct}
\eea
This result holds for $T >> T_{cross}$ where one can neglect
contributions to the conductivity coming from higher powers of
$t_{\perp}$. 

 The transverse optical conductivity is relatively easy to measure and
therefore one can use it to distinguish between Luttinger and Fermi
liquid states. We argue that $\s_{\perp}(\omega, T)$ has different
behavior as function of $\omega$ for $\theta > 1/2$ and $\theta <
1/2$. Indeed, as follows from Eq.(\ref{S}), in both cases the
conductivity is finite at $\omega = 0$. At $\omega >> T$ it becomes
independent on $T$: $\s_{\perp}(\omega >> T) \sim \omega^{- 1 +
2\theta}$. This means that the function $S(x)$ is finite at $x = 0$
and behaves  as $x^{- 1 + 2\theta}$ at $x >> 1$. At $\theta < 1/2$
this gives a peak at $\omega = 0$ and for $\theta > 1/2$ there is a
minimum. In the latter case $\s_{\perp}(\omega, T)$ is qualitatively
different from 
the coherent conductivity which has a Drude peak.

For $\theta < 1/2$ it is
more difficult because the
transverse conductivity acquires a maximum at $\omega = 0$ 
 which may be confused with the
Drude peak. In order to distinguish this incoherent maximum from the
Drude peak one should measure the temperature dependence of 
its width $\Delta_{\omega}$.  
According to Eq.(\ref{conduct}) $\Delta_{\omega} \sim T$; for a Fermi
liquid we expect $\Delta_{\omega} \sim T^2$.

 The Wen's criterion $\theta > 1$ is a sufficient one which is 
 guaranteed to work at $D
\rightarrow \infty$.  Now we are going to argue that at small $D \sim
2$ 
the incoherent Luttinger liquid  regime at finite temperatures 
may be realized
at sufficiently weaker conditions. At finite $D$ there is a
possibility that a 
three dimensional phase transition generated by the virtual interchain
processes will  happen at temperatures higher than those at which  
the single particle pole is formed. 
A probability of such outcome increases when $D$ decreases. The large
$D$ approximation  used in derivation of  Eq.(1)
becomes poor at small $D$, 
which opens an intriguing possibility that there is a critical value
of  $D$ below which a phase transition occurs at temperatures where
the Fermi liquid pole is not yet developed. This may explain why the 
effective reduction of dimensionality in (TMTSF)$_2$PF$_6$ caused by
a magnetic field has lead to non-Fermi-liquid effects 
reported  in Ref.\cite{chaikin}.  

 One may gain an insight  from studying a case of two chains. It was
pointed by Yakovenko \cite{yak} and Nersesyan {\it et al.}
\cite{kus}, who considered the case of spinless fermions,  
that under renormalization the  hopping generates an effective
interchain backscattering which (for two chains) may open a spectral
gap. It is plausible that for an infinite number of chains instead of
the gap one will get a phase transition. 
The corresponding renormalization group (RG) equations for fermions 
with spin were derived by Khveshchenko and Rice
\cite{khv}. Below we shall analyze these equations following the
approach of
Ref.\cite{yak}.

 Without a loss  of generality we may consider the case $K_c < 1$.
In this case the RG equations of  Khveshchenko and Rice
describe  three possible regimes.

(i) {\bf Strong confinement regime}. $K_c < 3 - 2\sqrt 2 \approx
0.17$. This corresponds to the regime $\theta > 1$. 

(ii) {\bf Weak  confinement regime}. $3 - 2\sqrt 2 < K_c < \sqrt 5 - 2
\approx 0. 24$. In this region, as we shall
demonstrate in a moment, both single particle tunneling $t_{\perp}$
and dynamically generated interchain backscattering grow, but the
latter interaction reaches
the strong coupling regime first.

(iii) {\bf Band theory regime.} $K_c > \sqrt 5 - 2$. In this regime the
best approach is to  diagonalize the quadratic fermionic  Hamiltonian
first  to obtain two split Fermi surfaces, and then take into account the
interactions.

 Let us consider the most interesting case (ii) which corresponds to 
\be
1/K_c - K_c > 4 \label{Kcr}
\ee
 It can be shown that at these values
of $K_c$ the effective exchange interaction is formed already at 
small distances. Condition (\ref{Kcr}) corresponds 
to rather small $K_c < 0.24$, where the
operators containing the dual fields $\Theta_{1,2}$ are  strongly
irrelevant.  In this case we can simplify the RG equations
Khveshchenko and Rice (see Eqs.(2. 6 - 21) in their paper)
by  retaining
only the coupling constants $g_1, g_5$ and $t_{\perp}$. Treating these
coupling constants as small we shall also
neglect   renormalization of
$K_{c,s}$ from their 
initial values $K_c < 1, K_s = 1$. Then we obtain the simplified version of 
the RG  equations derived in
Ref. \cite{khv}, namely, we shall consider the case $K_c < 1, K_s = 1$ and
neglect renormalization of these parameters. In this case Eqs.(2. 6 -
21) from Ref. \cite{khv} simplify:
\bea
\rd z / \rd l &=& \left[3 - \frac{1}{2}(K_c + 1/K_c)\right] z, \label{1} \\
\rd g / \rd l &=& (1 - K_c) g + (K_c - 1/K_c) z , \label{2}
\eea
where $g = g_1, g_5$ and $z = [t_{\perp}/\epsilon_{\rF}]^2$in the notations of
Ref. \cite{khv}.   We shall consider the case where the 
interchain exchange is  absent in the bare Hamiltonian and 
generated dynamically 
in the process of
renormalization. This corresponds to the following initial conditions:
$g(0) = 0$.

 In this approximation we
can solve Eq.(\ref{1}):
\be
z_l = z_0\exp\{[3 - (K_c + 1/K_c)/2]l\} \label{zfl}
\ee
Substituting this into Eq.(\ref{2}) and taking into account the
initial condition $g(0) = 0$ we get 
\be
g_l = - \frac{z_0(1/K_c - K_c)}{1/K_c - K_c
 - 4}\left\{1 - \exp[- l(1/K_c - K_c - 4)/2]\right\}\exp[(1 - K_c)l]
\ee
Let us suppose that $(K_c^{-1} - K_c - 4)$ is not very small. Then $g_l$
becomes of order of one at 
\be
l^* \approx - \frac{1}{(1 - K_c)}\ln z_0
\ee
Substituting it into Eq.(\ref{zfl}) we get 
\be
z(l^*) \approx z_0^{\frac{K_c^{-1} - K_c - 4}{2(1 - K_c)}} \label{condit}
\ee
Since the exponent is positive and $z_0 << 1$, this means that at 
the point where the effective  exchange is 
already strong and presumably opens a spectral gap, the single
particle tunneling is still weak. 

 As we have said, it is plausible that the gap 
opening in the two chain model is a
precursor of a phase transition in the two dimensional array of
chains. Thus we conclude that for $D = 2$ the conditions for existence
of Luttinger liquid 
are relaxed: instead of having Luttinger liquid phase 
at $K_c < 0.17$ one {\it may}
have it  (at finite temperatures, of course)
already at $K_c < 0. 24$. A presence of bare interchain interactions
(the Coulomb interaction, for instance) will further relax the condition
on $K_c$.

I am  grateful to L. Degiorgi for information about his experiments,
to N. Andrei, A. G. Green and V. Yakovenko 
 for valuable discussions and constructive criticism. I am
greatly indebted to K. Sch\"onhammer who pointed out the mistake in the
first version of the paper.

\newpage

\begin{figure}[htbp]
\centerline{\psfig{figure=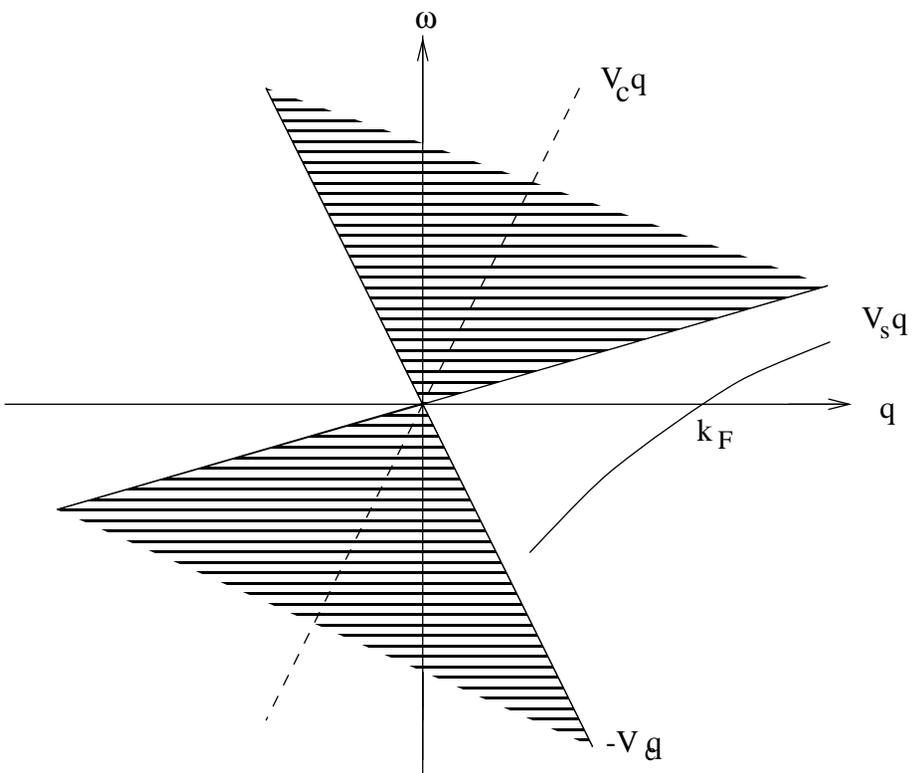,width=4in}}
\caption{The shaded region is the region of $(\omega, q)$ plane where 
$\Im m G_0^{(R)}$ for the single chain does not vanish. 
The solid line shows the dispersion law
of the coherent fermion excitation  near the right Fermi point.}
\label{fig.1}
\end{figure}

\begin{figure}[htbp]
\centerline{\psfig{figure=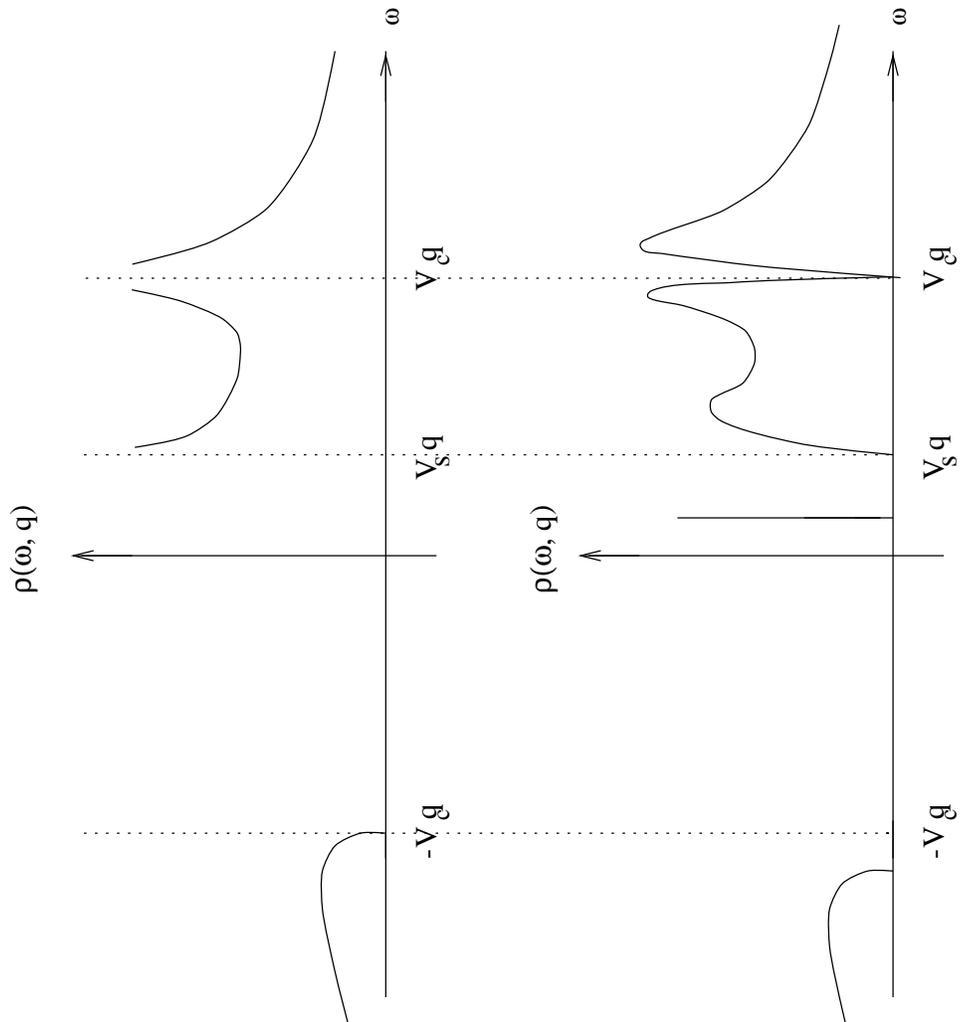,width=5in}}
\caption{(a) The spectral function 
for the spin-1/2 Luttinger liquid at $\theta < 1/2$, 
$q > 0$, (b) the same in the
presence of the quasiparticle peak. }
\label{fig.2}
\end{figure}

\end{document}